\documentstyle[11pt,paspconf]{article}
\begin{document}

\title{Nuclear kpc-sized disks of spiral galaxies}
\author{A.V.Zasov and A.V.Moiseev}
\affil{Stenberg Astronomical Institute,13 Universitetskij
Prospect, Moscow 119899, Russia}

\begin{abstract}
A complex structure of nuclear disks of normal spiral galaxies is
illustrated on the example of five galaxies, observed at 6m
telescope. A problem of gravitational stability of nuclear disks
is shortly discussed.

\end{abstract}

\section{Introduction}
Circumnuclear regions within a radius of a few hundred parsecs
are so diverse by their properties, that they may be called the
most unlike parts of galaxies. Usually these regions cannot
be considered as a simple continuation of the main disks of
parent galaxies, being decoupled by their structure, angular
velocity, gas content, star formation rate, or metal abundance.

In this paper we will touch upon two topics: structure of nuclear
disks, and the influence of their rotation on star formation.

\section{ Photometrical and kinematical structures of nuclear
 disks}

Fast rotating central parts of the observed galaxies as a rule do not
exceed 5-15\arcsec~by size, which makes their spectral study
a rather difficult task. It appears, that the reliable data for
their kinematic properties can't be obtained from one-dimensional spectral
cuts, actually two-dimensional velocity fields are necessary to distinguish
between circular and non-circular motions of gas or stars.

To illustrate the structural and dynamical properties
of the nuclear disks, we will discuss briefly the observational data
for the inner regions of five normal spiral galaxies: NGC 972, 1084,
4100, 6181 and 7217. All spectral observations were carried out at 6m
telescope of Special Astrophysical observatory in Russia (SAO RAS) in
1993 -1997. Scanning interferometer Fabry-Perot
(IFP) and Multi-Pupil Field Spectrograph   (MPFS) developed in this
observatory were used for obtaining the velocity field of the ionized gas
in the $\mbox{H}_\alpha$ emission line. Although the observations were
aimed mostly to study gas motions in a global galactic scale (program
"Vortex" led by Fridman A.), five objects chosen here may give bright
examples of different structures and kinematic peculiarities, observed
in nuclear regions of non-AGN galaxies.

The main results of observations are illustrated in Table 1.

   As Table shows, all galaxies have a circumnuclear
structure, containing at least two kinematic subsystems: the inner
one, which may be described as minibar or dynamically decoupled
disk, and the outer subsystem -- a co-planar or moderately
inclined gaseous disk, which contains  spiral or ring structures.

The common feature of many galaxies is the turn of kinematic major axes
in their circumnuclear region. There may be two probable interpretation
of this effect -- bar-like perturbation of the velocity field, and the
inclination of nuclear disk with respect to the the main disk.
To distinguish between these possibilities
photometrical data were involved in addition to kinematic ones.

A  general approach is quite simple. In the first case (the presence of
a bar) one can expect some characteristic distortion of velocity field
of gas. Both the models of gas dynamics in a barred potential
and the observations of barred galaxies show that isovelocity contours
in the vicinity of a bar tend to turn along the major bar axis,
so the line of largest velocity gradient (kinematic major axis)
turns towards the minor photometrical bar axis. Hence,
photometrically and dynamically obtained position angles (PAs) of
the major axes change in a different way, turning in the opposite
directions.
In the second case, when the plane of the circular rotating
nuclear disk is not coplanar to the main disk, the  photometric
axis always turns parallel to the dynamical one. Observations
show that both cases occur in real galaxies.

\begin{table}
\caption{Kinematical structures and features of nuclear disks}
\label{tlb}
\begin{center}
\scriptsize
\begin{tabular}{rlrrll}
NGC  & Type  &    R(\arcsec) & R(kpc)  &  Structure of nuclear region& Reference \\
\tableline
972 &  Sb   &      6       &   0.6   & minibar and pseudoring &Zasov \& Sil'chenko (1996)\\
        &       &     20       &   1.8   & star-forming ring + IR
        spiral (?) & Zasov \& Moiseev (1998) \\
1084&  Sc   &      6       &   0.6   & radial motion of the ionized gas &  Moiseev et al. in prep. \\
        &       &      20      &   1.8   & dynamically revealed bar   \\
4100&  Sbc  &      6       &0.4      &  blue optical ring &  Moiseev et al. in prep. \\
        &       &      11      &   0.9   &  inclined or polar star-forming disk  & \\
6181&  Sc   &      5       &   0.8   & minibar     & Sil'chenko et al. (1997)\\
        &       &      12      &   2.0   &co-planar disk+expanding ring& \\
        &       &              &        & or strong vertical  motion & \\
7217&  Sb   &       2      &   0.15  &  polar  disk and ring & Zasov \& Sil'chenko 1997 \\
        &       &      10    & 0.8&   fine  spiral  arms in co-planar disk \\
\end{tabular}
\end{center}
\end{table}

Some comments to the chosen galaxies:

{\bf NGC~972}. This is the isolated galaxy of unusual optical
structure with dusty disk and negligible bulge. Wide dust lane,
bordering the bright inner part of the galaxy in the south-west
side, gives some hint of the weak inclined inner disk of interstellar
matter which was confirmed by measuring the orientation of
line-of nodes. The galaxy is rich of molecular gas,
nevertheless it possesses
a rather moderate star formation which is concentrated in the
small nucleus and in the ring of about 2 kpc radius. The latter
is not visible in the images obtained at the 1m telescope SAO in
VRI bands, but clearly noticeable in $\mbox{H}_\alpha$ line and
also in the map of distribution of Q-parameter , which is absorption
-independent combination of V,R and I magnitudes (Zasov \&
Moiseev 1998). Q-parameter map also  shows a small pseudoring
with radius about $0.5\pm1$ kpc, which coincides with
$\mbox{H}_\alpha$ nuclear ring, found recently by Ravindranath \&
Prabnu (1998).

IFP velocity field of the inner region of about 0.6 kpc radius
reveals the turn of kinematical axis at about 30\deg  ~relative
to the orientation of the outer line-of-nodes found earlier from
the observations with  MPFS (Zasov \& Sil'chenko 1996). The
K-band image (which was obtained in UKIRT and kindly given by
Stuard Ryder to authors) shows that the isophotes major axes turn
in the opposite direction.  It gives evidence of minibar inside
the nuclear ring. In between the central disk and the ring the
rotation of gas is circular. The irregular structure of this
region, seen in K-band, resembles  widely opened spiral. So there
are two kinematic subsystems, coexisting in the inner part of
this galaxy.

{\bf NGC~1084.} This is a late-type galaxy with a well defined spiral
structure. Velocity field analysis of the very inner bright region
within 6\arcsec~ (0.6 Kpc) from the centre demonstrates a fast  circular rotation
of gas in the plane of the main disk, in addition to a less intense
shifted components of $\mbox{H}_\alpha$ and $\mbox{[NII]}$ line profiles,
revealing non-circular (probably, radial) motions. The most unusual
peculiarity of gas kinematics which was found from  IFP observations
is the long (about 1.5 kpc size) shock front, crossing the inner
part of the galaxy, where the line-of-sight velocity changes at
80-100 km/s within several arcsec perpendicular to the front
line. Direct images of  NGC 1084 were obtained at the 1m
telescope SAO in B,V,R,I bands.  Image processing showed the turn
of photometric major axis which has the opposite sign with
respect to the turn of kinematic one. It agrees well with the
presence of bar-like perturbation  with radius about 2 kpc. Why
the presence of a any optical feature in this region is not seen
in the image of this galaxy -- is a puzzle. It seems that the
contrast of the bar potential is rather weak.

{\bf NGC~4100}. This is non-barred galaxy in Ursa Major cluster. Its
fast rotating nucleus was found by Afanasiev et al.(1992).
 Analysis of
the velocity field showed, that in the  inner 11\arcsec, or about
0.8 kpc, where a very large velocity radial gradient is present, a
kinematical axis  turns  by 20\deg ~with respect to the outer disk.
Ellipticity of the isophotes
is maximal at the central 6\arcsec. The  photometric axis
in R,I bands turns parallel  to the kinematical one, which indicates
the existence of the inclined inner disk in the central 0.8-0.9 kpc.
In the frame of circular gas motions  the dynamically determined
inclination of the main disk and the  nuclear region differ by 22
\deg ~ in such a way that the nuclear disk looks more
opened for the observer. There are two possible solutions for the
inclination angle between the planes of two disks: 25\deg~ or
87\deg, ~depending on what side of nuclear disk is closer to us.
Blue color index and bright $\mbox{H}_\alpha$ emission gives
evidence of intense SF in the nuclear disk.

{\bf NGC~6181}.
In addition to IFP observations, the images of this
SAB(rs)c galaxy, obtained at 1m telescope, were used (see Sil'chenko et al.
1997). In the very centre of this galaxy, at radius of about 0.6-0.8
~kpc, both kinematic and photometric major axes turn in the opposite
directions, which enables to conclude that a small nuclear bar exists
in this galaxy.

The central part of the residual velocity field, obtained by the
subtraction of the observed and the expected circular
velocity fields, reveals two
ring-like arcs at radius 1.8 kpc, simmetrically positioned near
the minor axis, where deviations from the circular rotation
locally exceed 50 km/s. It is not clear what may cause such
strong radial motions (or z-motions) of gas -- these regions do
not reveal themselves in brightness distribution. There is also
no hint of the related shock wave or the enhanced line emission.
A plausible explanation is that we observe here an unusually
large amplitude of oscillation of gas velocities associated with
the density waves, which penetrate deep into the inner part of the
disk.

{\bf NGC~7217}. Contrary to the galaxies, discussed above, this galaxy
possesses a high luminous spherical component. Although this galaxy
has two or three optical rings, there is no bar at neither
visible, nor near-infrared wavelengths. Observations with  IFP
and  MPFS show that the azimuthal variation of the
line-of-sight velocity gradient follows a nearly sinusoidal
curve, which indicates that the gas moves along circular orbits.
In the circumnuclrear region this gradient amplitude sharply
increases up to $250~\mbox{km}/\mbox{s}/\mbox{kpc}$. The
significant turn of kinematic line-of nodes, is also noticeable.
So, in this galaxy we have an example of sharply kinematic distinct nucleus.

Our data show a rapid decrease in the photometric PA of the major axis
toward the center of the galaxy, beginning from about 4\arcsec. However,
the HST measurements, taken from the NASA/ESA archive data, show that
this decreasing actually occurs closer to the center
-- at a distance of about 1\arcsec ~in such a way, that the central
isophotes become nearly perpendicular to the outer ones. They
also increase their ellipticity toward the centre. The kinematic
axis orientation is in satisfactory agreement with the
photometric estimates. Therefore, the most likely explanation for
the rotation of the photometric and dynamical axes is the
presence of a small strongly inclined, (probably, polar) disk in
central 100-200 pc region of the galaxy.
\footnote {Small polar nuclear disks in normal spiral galaxies
probably are  not so seldom: for example,  their presence was
claimed in NGC~2685 (Sil'chenko et al. 1998), NGC~253
(Ananthramaiah \& Goss 1996) and some other galaxies.}

\section{Nuclear disk stability and star formation}

A fast rotation of nuclear disks of galaxies is a factor, which
tends to reduce the star formation activity due to angular
momentum of collapsing gas regions, which prevents gaseous disks
from being gravitationally unstable.

A flat gaseous disk of the surface density $\sigma_{gas}(R)$ is
gravitationally stable
if radial velocity dispersion of gas $C_{gas}$ is high enough for
the Toomre $Q$-parameter ($Q \sim C_{gas}\cdot \kappa (R) /
\sigma_{gas}(R)$, where $\kappa (R)$ is the epicyclic frequency)
to be larger than some critical value
$Q_c$, so that $Q_c=1$ for pure radial perturbations (Toomre'
criterion). In disks of spiral galaxies $Q_c=1.5-2$
(Kennicutt 1989, Zasov \& Bizyaev 1994). Non-WKB analysis of
stability  shows that the threshold for instability $Q_c \approx
1.7$ for 'flat' rotation curve, but keeps close to 1 if the
angular velocity $ \Omega
\approx \mbox{const}$ (Polyachenko et.al. 1997). It follows then, that in the
case of rigid-body rotation, which usually takes place in a
circumnuclear regions, a higher value of the gas surface density
$\sigma_{gas}$ is necessary for the disk to be unstable -- due to lower
$Q_c$ and higher $\kappa(R)$.

Indeed, many spiral galaxies possess dense molecular circumnuclear disks of
about one kpc size, for which $\sigma_{gas}$ exceeds
$10^3\,M_\odot / pc^2$, so a
large angular velocity is necessary to stabilize the disk.
However, as a rule, values of  $ \kappa (R)$ for them
are also very high, and, as a result, the velocity dispersions $C_{gas}$,
corresponding to $Q_c \approx 1$, remain rather low.
The estimates of marginal values of
$C_{gas}$ (from the data taken in the literature) for about two
tens of galaxies which have molecular nuclear disks, shows, that
for most of them $C_{gas} \le 15 \mbox{km}/\mbox{s}$ which does not exceed the
observed velocity dispersion of gas (Zasov 1999). This
result gives evidence that in many cases nuclear molecular disks are on the
threshold of gravitational stability or definitely stable. The
latter is is especially true for nuclear regions of galaxies poor
of gas, such as NGC~7217. Nevertheless some star
formation takes place even there. In NGC~7217 not only
the growth of the  intensity of $\mbox{H}_\alpha$ towards the
centre is observed, but also, as the analysis of HST observations
showed, a surprisingly well-ordered spiral-like structure exists
within inner 10\arcsec~ (Zasov \& Sil'chenko 1997).

Even if the inner disk of NGC~7217 is marginally stable, the
wavelength of growing gravitational perturbations is
expected to be of about several hundreds of parsecs there,
whereas the observed structure presents a sort of "rippled
surface" with a significantly smaller scale. It confirms that the
observed pattern cannot be caused by gravitational oscillations.

Note that the small-scale spiral pattern may frequently occur in
the nuclear disks, although it is usually difficult to extract it
from the photometrical observations restricted angular
resolution. As an example, a complex circumnuclear spiral-like
structure was found in NGC 6951 (Barth et al. 1995) and NGC 488
(Sil'chenko 1999).

A possible alternative mechanism of formation of spiral pattern
is the hydrodynamical instability in the gaseous disks which does
not require a high surface density to develop (see the
discussion by Fridman (1994)). So the presence of star formation
in rapidly rotating disks may give evidence of the importance of
non-gravitational mechanism of compression of gaseous medium there.

\acknowledgments{ Authors are  grateful to   Afanasiev V., Boulesteix J., Burenkov A., Dodonov S.,
 Sil'chenko O. and Vlasyuk V. for the obtaining the observational
 data. This work was supported by grant RFBR 98-02-17102}.

\end{document}